\documentclass[twocolumn,showpacs,prb]{revtex4}% Physical Review B
\usepackage{bm}
\usepackage{graphicx}
\newcommand{\rot}{\mathop{\mathrm {rot}}\nolimits}

\newcommand{\nix}[1]{}

\begin{document}

\title{Resonant three-dimensional photonic crystals}
\author{E.L.~Ivchenko}
\author{A.N.~Poddubny}
\affiliation{A.F.~Ioffe Physico-Technical Institute, Russian
Academy of Sciences, St. Petersburg 194021, Russia}
\begin{abstract}
We have developed a theory of exciton-polariton band structure of
resonant three-dimensional photonic crystals for arbitrary
dielectric contrast and effective mass of the exciton that is
excited in one of the compositional materials. The calculation has
been carried out for a periodic array of spheres embedded in a
dielectric matrix. It has been shown that the position of the
lower branches of the polariton dispersion curve monotonously
depends on the exciton effective mass and is determined by the
coupling of light with the first few states of the mechanical
exciton quantum confined inside each sphere. Particularly, we have
studied the role of excitonic effects on the photonic-crystal band
gap along the $[001]$ direction of the Brillouin zone and
presented an analytic description of the polariton dispersion in
terms of the two-wave approximation.
\end{abstract}

%73.21.Fg (electron states and collective excitations in multilayers
%QW's, mesoscopic, and nanoscale systems [quantum wells])
%78.67.De (optical properties of nanoscale materials and structures
%[quantum wells]).
\maketitle

\section{Introduction}
The concept of photonic crystals was put forward by Yablonovich
\cite{yablonovich} and John \cite{john} in 1987. Since then this
term is used for media with the dielectric susceptibility varying
periodically in the space and allowing the Bragg diffraction of
the light. The theory of photonic crystals has been developed in a
number of subsequent works, see, e.g.,
Refs.~\onlinecite{soukoulis,meade,hornreich,sipe,busch}. The main
goal of these studies was to determine the photon band structure
and analyze the sequence of allowed bands and stop-bands (gaps)
for different directions of the wave vector in the first Brillouin
zone. The simplest model realization of a photonic crystal is a
structure grown from two materials, A and B, with different
dielectric constants $\varepsilon_{A}$ and $\varepsilon_{B}$: a
periodic layered medium ...A/B/A/B... in case of {\it
one-dimensional} photonic crystals and periodic arrays of
cylinders and spheres of the material A embedded in a dielectric
matrix B, in case of {\it two-dimensional} and {\it
three-dimensional} photonic crystals, respectively. Periodic
structures where the dielectric susceptibility of one of the
compositional materials, as a function of the frequency $\omega$,
has a pole at a certain resonance frequency are grouped into a
special class of resonant photonic crystals. In such systems the
normal light waves are excitonic polaritons.
%''If a three-dimensionally periodic structure has an
%electromagnetic band gap which overlaps the electronic band gap,
%then spontaneous emission can be rigorously
%forbidden.''\cite{yablonovich}
%''Therefore it would be very interesting to create periodic
%three-dimensional dielectric structures in which there exists an
%electromagnetic band gap.''\cite{yablonovich}
%``It has been shown that strong localization of photons may occur
%in a highly predictable manner in a frequency window in certain
%disordered superlattice microstructures of sufficiently high
%dielectric contrast.''\cite{john}
In Refs.~\onlinecite{john1,phonon} the dispersion of light waves
has been calculated taking into consideration the frequency
dependence of the dielectric susceptibility in the frame of the
local material relation ${\bm D} = \varepsilon_A(\omega) {\bm E}$
between the electric displacement and the electric field. In
Ref.~\onlinecite{nocontrast} the dispersion of excitonic
polaritons in a resonant photonic crystal has been calculated
taking into account only one level of the quantum-confined exciton
in a sphere A and neglecting the difference between the dielectric
constant $\varepsilon_B$ and the background dielectric constant
$\varepsilon_a$ of the material A. In the present work we have
studied theoretically the dispersion of excitonic polaritons in a
resonant photonic crystal making allowance for all available
exciton quantum-confinement levels as well as for the dielectric
contrast, i.e., for $\varepsilon_a \neq \varepsilon_B$.
\section{Problem definition and method of calculation}
In this work the theory of resonant photonic crystals is developed
for a periodic array of spheres A arranged in a face-centered
cubic (FCC) lattice and embedded in the matrix B. The structure
under consideration is characterized by seven parameters: $R$,
$a$, $\varepsilon_B$, $\varepsilon_a$, $\omega_{\rm LT}$,
$\omega_0$ and $M$. Here $R$ is the radius of the spheres A, $a$
is the lattice constant for the FCC lattice, $\varepsilon_B$ is
the dielectric constant of the matrix, $\omega_0$, $\omega_{\rm
LT}$ and $M$ are the resonance frequency, longitudinal-transverse
splitting and translational effective mass of the triplet $1s$
exciton excited inside the spheres A, $\varepsilon_a$ is the
background dielectric constant that contains contributions to the
dielectric response from all other electron-hole excitations.
Thus, we assume the dielectric function of the bulk material A to
have the form
\begin{equation} \label{dielect}
\varepsilon_A(\omega, q) = \varepsilon_a + \frac{\varepsilon_a
\omega_{\rm LT}}{\omega_{\rm exc}(q) - \omega}\:,\: \omega_{\rm
exc}(q)=\omega_0+\frac{\hbar q^2}{2M}
\end{equation}
and, in that way, take into account both the frequency and spatial
dispersion, i.e. the dependence on the light frequency $\omega$
and wave vector ${\bm q}$. The radius $R$ is chosen so that, on
the one hand, the spheres A do not overlap, i.e. $R <
a/2\sqrt{2}$, but, on the other hand, $R$ should exceed the Bohr
radius of the $1s$-exciton in the material A and, hence, the
exciton can be considered as a single particle with the mass $M$.
In the following we ignore the frequency dependence of the
parameters $\varepsilon_B$ and $\varepsilon_a$. Moreover,
hereafter we assume the material A to be isotropic and take into
consideration the bulk $1s$ exciton states only. It follows then
that the problem is reduced to the solution of a system of two
vector equations, namely, the wave equation
\begin{equation} \label{waveeq}
\rot\rot{\bm E}({\bm r}) = \left( \frac{\omega}{c} \right)^2
\bigl[ \varepsilon({\bm r}){\bm E}({\bm r}) + 4 \pi {\bm P}_{\rm
exc}({\bm r}) \bigr]
\end{equation}
and the material equation for the $1s$-exciton contribution to the
dielectric polarization
\begin{equation} \label{mecheq}
\left( -\frac{\hbar}{2M}\ \Delta + \omega_0 - \omega \right) {\bm
P}_{\rm exc}({\bm r})=\frac{\varepsilon_{a}\omega_{\rm LT}}{4\pi}
{\bm E}({\bm r})\:,
\end{equation}
where $\varepsilon({\bm r}) = \varepsilon_a$ inside the spheres
and $\varepsilon({\bm r}) = \varepsilon_B$ outside the spheres,
${\bm E}({\bm r})$ and  ${\bm P}_{\rm exc}({\bm r})$ are the
electric field and the excitonic polarization at the frequency
$\omega$. On a spherical surface separating materials A and B we
impose the standard Maxwell boundary conditions: continuity of the
tangential components of the electric and magnetic fields, and the
Pekar additional boundary condition for the excitonic
polarization: the vanishing vector ${\bm P}_{\rm exc}({\bm r})$ at
$|{\bm r} - {\bm a}| = R$, where the translational vectors ${\bm
a}$ define the centers of spheres A.

Due to the periodicity of the structure we can seek solutions of
Eqs.~(\ref{waveeq}) and (\ref{mecheq}) in the Bloch form
satisfying the condition
\begin{equation}\label{bloch}
{\bm E}_{\bm k}( \bm{ r + a})={\rm e}^{{\rm i} {\bm k}{\bm a}}
{\bm E}_{\bm k}(\bm r)\:.
\end{equation}
Here ${\bm k}$ is the exciton-polariton wave vector defined within
the first Brillouin zone. We remind that, for a FCC lattice, the
latter is a dodecahedron bounded by six squares and eight
hexagons.

Below we present the results of calculation of the
exciton-polariton dispersion $\omega_{n {\bm k}}$, where $n$ is
the branch index. The computation was mainly performed by using a
photon analogue of the Korringa-Kohn-Rostoker (KKR) method
\cite{kkr,moroz,moroz1}. In this method (i) the electric field is
decomposed in the spherical waves, or more precisely, in the
vector spherical functions centered at the points ${\bm r} = {\bm
a}$, and (ii) following the consideration of the light scattering
by a single sphere and the introduction of a structural factor the
dispersion equation is transformed to
\begin{equation} \label{det_disp}
\left\vert \delta_{j^{\prime} j} \delta_{m^{\prime} m}
\delta_{\sigma^{\prime} \sigma}- G_{j^{\prime} m^{\prime}
\sigma^{\prime} , j m \sigma}({\bm k},\omega) R_{j
\sigma}(\omega) \right\vert = 0\:.
\end{equation}
Here $R_{j \sigma}$ are the coefficients describing the scattering
of spherical waves by a single sphere A, they depend on the total
angular momentum $j$ and the polarization index $\sigma$
discriminating the magnetic and electric spherical harmonics but
are independent of the angular-momentum component $m$. Note that,
for a spherical scatterer, these coefficients relate the incident
field ${\bm E}_0({\bm r}) \propto {\bm J}_{j m \sigma}({\bm r})$
with the scattered field ${\bm E}_{\rm sc}({\bm r}) \propto {\bm
H}_{j m \sigma}({\bm r})$, where ${\bm J}_{j m \sigma}, {\bm H}_{j
m \sigma}$ are the vector spherical functions \cite{kkr}. Ajiki et
al.~\cite{kcho} have calculated values of $R_{j \sigma}(\omega)$
taking into account the exciton resonance in the spheres A and the
finite exciton effective mass $M$. In contrast to the scattering
matrix for a single sphere which is diagonal, $R_{j' m' \sigma', j
m \sigma} \equiv R_{j \sigma} \delta_{j' m' \sigma', j m \sigma}$,
the matrix of structural factors $G_{j^{\prime} m^{\prime}
\sigma^{\prime} , j m \sigma}({\bm k},\omega)$ has nonvanishing
diagonal and off-diagonal components. It should be mentioned that
the both matrices are frequency dependent whereas only the
structural matrix ${\bm G}$ depends upon the polariton wave vector
${\bm k}$. At the same time ${\bm G}$ is independent of the
excitonic parameters and coincides with the matrix considered in
Refs.~\onlinecite{kkr,moroz,moroz1} where the exciton states are
disregarded.

In addition to the KKR method, in section 4 we apply the
Green-function technique for the analysis of separate
contributions of quantum-confined excitonic levels to the
polariton dispersion, and in section 5 we use the two-wave
approximation which allows an analytic description.
\section{Polariton dispersion for a finite exciton
mass} First we focus the attention upon purely excitonic effects
and neglect the dielectric contrast assuming $\varepsilon_a =
\varepsilon_B$. Then in the absence of exciton-photon interaction,
i.e. for $\omega_{\rm LT} = 0$, the medium becomes optically
homogeneous and photons propagating therein are characterized by
the linear dispersion $\omega = c q /n_B$ with the refractive
index $n_B = \sqrt{\varepsilon_B}$. In the reduced zone scheme,
this single-valued relation between the frequency and wave vector
${\bm q}$ turns into the many-valued (or multi-branch) dispersion
curve
\begin{equation} \label{bare}
\omega_{\bm k} = c | {\bm k} + {\bm b}| /n_B\:,
\end{equation}
where ${\bm b}$ is the reciprocal lattice vector reducing ${\bm
q}$ to the vector ${\bm k} = {\bm q} - {\bm b}$ that lies in the
first Brillouin zone. A nonvanishing value of $\omega_{\rm LT}$
leads to a mixing between photonic and excitonic states and
formation of hybrid polariton excitations characterized by a
complicated multiband dispersion  $\omega_{n {\bm k}}$. As a
result the wave $(n, {\bm k})$ is a mixture of two or more
photonic states (\ref{bare}) with the same ${\bm k}$ but different
${\bm b}$.
\begin{figure}[t]
  \centering
    \includegraphics[width=.4\textwidth]{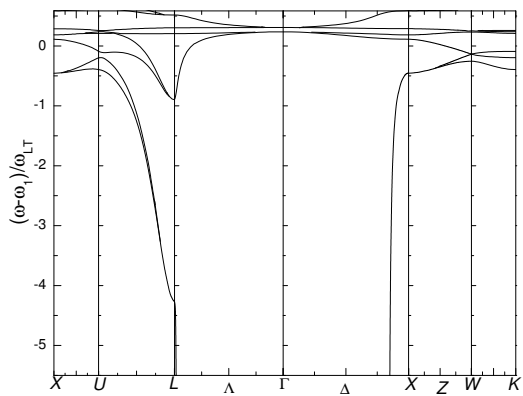}
  \caption{ Exciton-polariton band structure of a photonic crystal
  with a FCC lattice of spheres A inserted into the matrix B. The
  calculation is performed neglecting the dielectric contrast and
  for the set of parameters indicated in the text.
  }\label{pic1}
\end{figure}

Fig.~1 presents the dispersion of exciton-polaritons calculated
for a FCC lattice and the following set of structure parameters:
\[
\varepsilon_a = \varepsilon_B = 10\:,\: R = a/4 \:,\: \hbar
\omega_1 = 2\ \mbox{eV}\:,\: \omega_{\rm LT} = 5 \times 10^{-4}\
\omega_1\:,
\]
\[
P \equiv \left( \frac{\sqrt{3} \pi c}{\omega_1 n_B a}\right)^3 =
1.1 \:,\: M = 0.5\ m_0\:.
\]
Here $m_0$ is the free electron mass in vacuum and, instead of the
bulk exciton resonance frequency $\omega_0$, we introduced the
resonance frequency
\begin{equation} \label{1s1s}
\omega_1 = \omega_0 + \frac{\hbar}{2M} \left( \frac{\pi}{R}
\right)^2
\end{equation}
of the ground state of the exciton quantum-confined in a sphere of
the radius $R$. Note that, for the structure tuned to the Bragg
resonance $\omega_1 = c k_L/n_b$ at the point $L$ of the Brillouin
zone with $k_L = \sqrt{3} \pi/a$ or $\omega_1 = c k_X/n_b$ at the
point $X$ with $k_X = 2 \pi/a$, the parameter $P$ equals to 1 and
$3\sqrt{3}/8 \approx 0.65$, respectively. For $P = 1.1$, the
anticrossing between the horizontal line $\omega$$~=~$$\omega_1$
(``bare'' exciton branch) and the line $\omega = c k /n_B$
(``bare'' photon branch) occurs inside the Brillouin zone at $k$
amounting approximately 97\% of $k_L$ and 84\% of $k_X$. For the
sake of completeness  the dependence $\omega({\bm k})$ is shown
not only along the high-symmetry crystallographic directions ${\bm
k}
\parallel [001]$ (points $\Delta$) and ${\bm k}
\parallel [111]$ (points $\Lambda$) but also along the straight lines
$X-W$, $W-K$, $X-U$ and $U-L$.

In Fig.~1 the frequency region is cut from above in order to keep
only few branches of the dispersion curve. In the cut region the
dispersion is presented by a dense network of polariton branches
which appear as a result of the anticrossing of ``bare'' photonic
branches (\ref{bare}) with a set of close-lying exciton
quantum-confinement levels. The network has a tangled character
complicated for depiction. Therefore, we concentrate here on the
analysis of the role played by the photonic-crystal parameters in
the formation of the lower polariton branches.

In Fig.~2 the solid curves show the same dispersion branches as in
the previous figure but in an increased scale and in the vicinity
of the points $X$ and $L$. For comparison we also present, by
dashed-and-dotted lines, the lower branch of the dispersion curve
calculated taking into account only the ground exciton
quantum-confinement level. The calculation was performed by
reducing the value of $M$ to $0.01 m_0$ and decreasing $\omega_0$
so that the frequency $\omega_1$ in Eq.~(\ref{1s1s}) was kept
invariant. The dashed-and-dotted lines coincide with those
calculated by the method proposed for exciton-polaritons in Ref.~
\onlinecite{nocontrast}. Dashed curves illustrate the opposite
limiting case of quite heavy excitons, $M \rightarrow \infty$. In
this case the relation between the exciton polarization and
electric field becomes local:
$$
{\bm P}_{\rm exc} = \chi {\bm E}\:, \hspace{5 mm} \chi =
\frac{\varepsilon_a}{4 \pi} \frac{\omega_{\rm LT}}{\omega_0 -
\omega}\ .
$$
It means that values of ${\bm k}$ corresponding to a certain
frequency $\omega$ can be found in the same way as it is done for
nonresonant photonic crystal with the dielectric susceptibilities
$\varepsilon_B$ and $\varepsilon_A = \varepsilon_a + 4 \pi \chi$.
The calculation shows that the dashed line in Fig.~2 is
practically indistinguishable from the lower branch obtained
according to Eq.~(\ref{det_disp}) for $M = 5 m_0$. The lower
polariton branch is formed as a result of ``repulsion'' of the
photon branch (\ref{bare}) with ${\bm b} = 0$ towards the
long-wavelength side because of the interaction with the exciton
quantum-confinement levels. For $M \rightarrow 0$ but $\omega_1 =
{\rm const}$, this branch is remarkably affected by the lower
level (\ref{1s1s}) only. For $M \rightarrow \infty$, the other
levels act upon this branch to the maximum since in this limit
their resonance frequencies coincide and are equal to $\omega_0$.
We conclude then that the lower polariton branch corresponding to
the finite mass $M$ must always lie between the dashed-and-dotted
and dashed lines, in agreement with the results of calculation
shown in Fig.~2.
\begin{figure}
%[t]
  \centering
    \includegraphics[width=.4\textwidth]{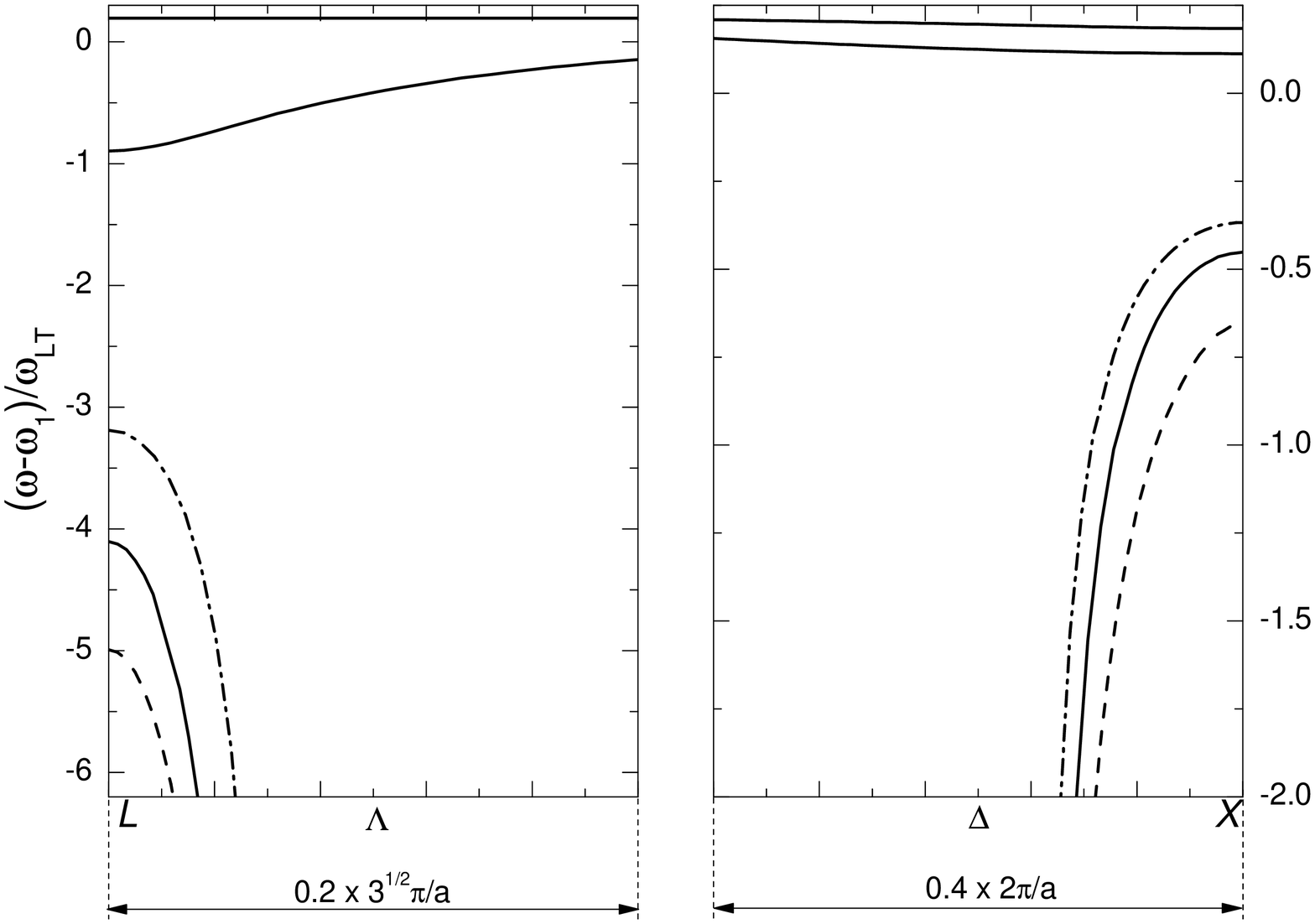}
  \caption{Dispersion of exciton polaritons in a resonant photonic
crystal in the spectral region adjoining the resonance frequency
of the lower exciton level $\omega_1$ and for the wave vectors
${\bm k} \parallel [111]$ (points $\Lambda$) and ${\bm k}
\parallel [001]$ (points $\Delta$). The dispersion curves are calculated
for the exciton effective mass $M = 0.5\,m_0$ (solid lines), $M
\rightarrow \infty$ (dashed lines) and $M \rightarrow 0$
(dashed-and-dotted lines). Other parameters are the same as in
Fig.~1.
  }\label{pic2}
\end{figure}
\section{Comparative contribution of individual
quantum-confined exciton levels} In the following we use the
notation $\omega_X(M)$ for the frequency at the $X$ point in the
lowest polariton branch for the exciton effective mass $M$ in the
material A, $\omega_X(0)$ and $\omega_X(\infty)$ for values of
this frequency for $M \rightarrow 0$ and $M \rightarrow \infty$,
and, finally, $\bar{\omega}_X$ for the arithmetic mean
$[\omega_X(0) + \omega_X(\infty)]/2$. One can see from Fig.~2 that
the difference $\omega_1 - \bar{\omega}_X$ noticeably exceeds the
difference $\omega_X(0) - \omega_X(\infty)$. This means that the
main contribution to the position of the frequency $\omega_X(M)$
should come from the exciton-photon coupling with the ground level
(\ref{1s1s}). In this section we analyze in detail the influence
of the ground and excited levels of the quantum-confined exciton
on the lower polariton branch. To this end we follow Ref.~
\onlinecite{kcho}, use the Green-function technique and expand a
solution of equation~(\ref{mecheq}) with arbitrary function ${\bm
E}({\bm r})$ in a series
\begin{widetext}
\begin{equation} \label{levelsum}
{\bm P}_{\rm exc}(\bm r) = \frac{\varepsilon_a}{4\pi}
\sum\limits_{\nu} \frac{\omega_{LT}}{\omega_{\nu} - \omega}
\sum\limits_{\bm a} \Phi_{\nu}({\bm r} - {\bm a})
\int\limits_{|{\bm r'} - {\bm a}| <R} \Phi_{\nu}^*({\bm r}' - {\bm
a})\bm E(\bm r') d^3r'
\end{equation}
\end{widetext}
over the eigen states of the mechanical exciton in a sphere with
an infinitely high barrier. Here the index $\nu = (n_r,l,m)$
characterizes the exciton state and comprises, respectively, the
radial quantum number, angular momentum and its projection on the
$z$ axis, $\omega_{\nu}$ is the exciton resonance frequency in the
state $\nu$. It is worth-while to note that the functions
$\Phi_{\nu}({\bm r})$ satisfy Eq.~(\ref{waveeq}) with
$\omega=\omega_{\nu}$ and without the inhomogeneous term, i.e.,
with ${\bm E} (\bm r)\equiv 0$, they are normalized to unity and
correspond to the sphere centered at the point ${\bm r} = 0$. The
frequencies $\omega_{\nu}$ are conveniently presented as $\omega_0
+ (\hbar/2MR^2) x_{n_r,l}^2$, where $x_{n_r,l}$ are dimensionless
numbers. For the few lowest energy levels, their values are given
by $\pi$ ($1s$), 4.493 ($1p$), 5.764 ($1d$), 2$\pi$ ($2s$), 6.988
($1f$), see e.g. Ref.~\onlinecite{zeros}, with symbols in
parenthesis indicating the radial number $n_r$ and the orbital
momentum, $s$ for $l=0$, $p$ for $l=1$ etc. Note that these
symbols characterize the exciton quantum-confined state whereas
the internal exciton state is $1s$ all the time. Substituting the
expansion (\ref{levelsum}) into the wave equation (\ref{waveeq})
and keeping in the sum over ${\nu}$ any one term or a finite
number of terms we can study the effect of these terms on the
formation of the polariton dispersion. The results are illustrated
in Fig.~3. It is seen that (a) the position of the frequency
$\omega_X(M)$ is mainly determined by the exciton level $1s$, (b)
the polariton frequency at the $X$ and $L$ points is unaffected by
the $1p$ level, and (c) it suffices to add the contributions from
the $1d$ and $2s$ levels in order to describe quite well the
position of $\omega_X(M)$ whereas, at the $L$ point, the sum
converges slowly. Obviously, this difference in the convergence is
related to the choice of $P=1.1$ in which case the effect of
anticrossing near the $L$ point is much stronger compared with
that at the $X$ point.

\begin{figure}
%[t]
  \centering
    \includegraphics[width=.4\textwidth]{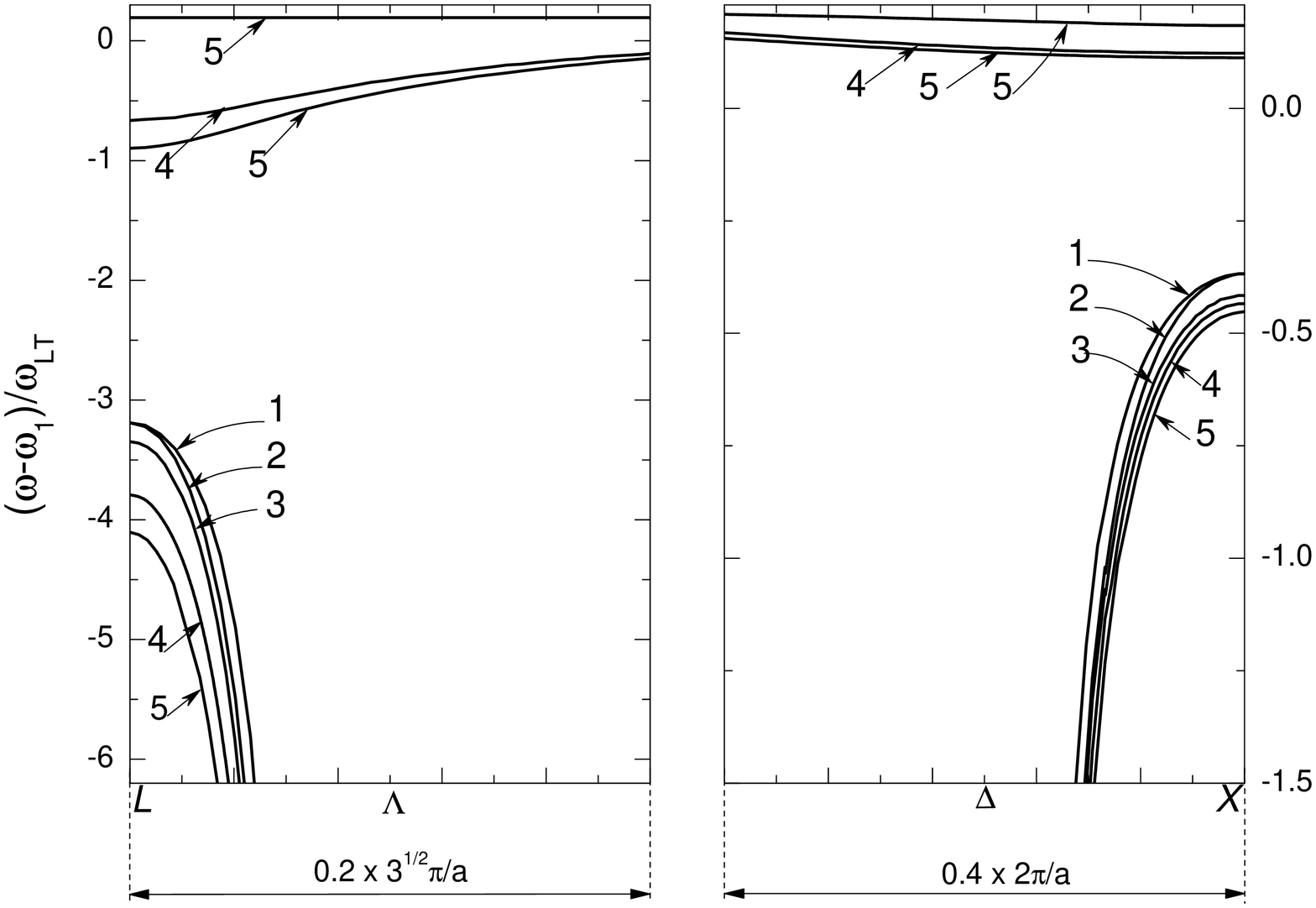}
  \caption{
Dependence of the exciton-polariton spectrum on the number of
excitonic levels taken into account while calculating the
dispersion curve by the Green-function technique. Lines 1, 2, 3
and 4 are obtained with allowance made, respectively, for one,
two, three and four lower levels; line 5 represents the exact
result taking into account all levels of exciton quantum
confinement. The structure parameters used are the same as in
Fig.~1.
  }\label{pic3}
\end{figure}
\section{Polaritons in photonic crystals with a dielectric contrast}
The dashed lines 1 and 2 in Fig.~4 depict the dispersion of light
waves in the vicinity of the $X$ point of the Brillouin zone in a
nonresonant photon crystal. Exactly at the $X$ point these waves
are characterized by the symmetry $X_{5}$ and $X_{5'}$. The solid
lines 1$'$, 2$'$ and 3$'$ show the dispersion branches calculated
taking into account only the exciton-photon coupling with one
excitonic level $1s$ and choosing its resonance frequency in the
middle between the frequencies of the ``bare'' photons $X_{5}$ and
$X_{5'}$. An additional branch 4$'$ represents longitudinal
exciton states which are practically dispersionless, in the
following it is not discussed. Finally, the dashed-and-dotted
lines 1$''$, 2$''$ illustrate the exciton-polariton dispersion
with allowance made for all exciton levels. For the illustration
we chose a photonic crystal with comparatively weak dielectric
contrast, $\varepsilon_B = 12$, $\varepsilon_a = 13$, in order to
have the lowest stop-band in the direction ${\bm k} \parallel
[001]$, or the splitting between the $X_{5}$ and $X_{5'}$ states,
comparable with the matrix element of exciton-photon interaction.
It follows from Fig.~4 that the allowance for the $1s$
quantum-confined exciton states optically active in the
polarization ${\bm E} \perp z$ results in a replacement of two
branches 1, 2 by three branches 1$'$, 2$'$ and 3$'$. The addition
of contributions due to other (excited) excitonic levels leads to
a transformation of the branch 1$'$ into 1$''$ and an appearance
an extra branch 2$''$. Simultaneously a dense network of polariton
branches is formed in the frequency region $\omega > \omega_0$. It
is not shown in the figure in order to avoid overloading the
image. Instead, only the exciton-polariton eigen frequencies
$\omega_{\nu} > \omega_0$ at the $X$ point are indicated by short
horizontal lines intersecting the vertical line $X$.
\begin{figure}
%[t]
  \centering
    \includegraphics[width=.4\textwidth]{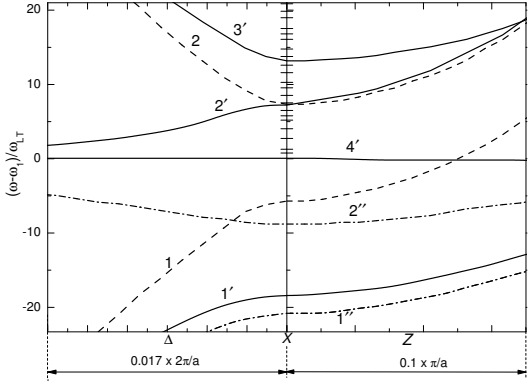}
  \caption{Photon band structure for a crystal with the dielectric
contrast, $\varepsilon_a = 12,\: \varepsilon_B = 13$. Dashed lines
are calculated for a photonic crystal neglecting excitons; solid
lines are obtained taking into account only the lowest excitonic
level; dashed-and-dotted lines and short straight lines are
obtained taking into account all quantum-confined states of the
mechanical exciton.
  }\label{pic4}
\end{figure}

In what follows we propose a two-wave approximate description of
the polariton spectrum which is valid for a weak dielectric
contrast and allows to understand the nature of branches indicated
in Fig.~4 by primed and double-primed integers. Earlier this
approximation was used in the analysis of the photon dispersion in
nonresonant photonic crystals \cite{satpathy}.

The Bloch solution (\ref{bloch}) can be expanded over the space
harmonics with the wave vectors ${\bm k} + {\bm b}$. For the
vectors ${\bm k} \parallel [001]$ lying close to the $X$ point, we
keep in this expansion only two terms with ${\bm k}_1 =
(0,0,k_1)$, ${\bm k}_2 = (0,0,k_2)$ and $k_1 - k_2 = 4\pi/a$. At
the $X$ point we have $k_1 = - k_2 = 2 \pi /a$. Thus, the electric
field is approximated by
\begin{equation} \label{twowaves}
{\bm E}({\bm r}) = {\bm E}_1 {\rm e}^{{\rm i} k_1 z} + {\bm E}_2 {\rm
e}^{{\rm i} k_2 z}\:.
\end{equation}
For doubly-degenerate polariton states of the $\Delta_5$ symmetry
compatible with the representations $X_5$ and $X_{5'}$, the
amplitudes ${\bm E}_1$ and ${\bm E}_2$ are parallel to each other
and perpendicular to the axis $z \parallel [001]$. Substituting
Eqs.~(\ref{levelsum}) and (\ref{twowaves}) into the wave equation
(\ref{waveeq}), multiplying its terms by $\exp{(- {\rm i} k_j z)}$
and integrating over the unit primitive cell $v_0 = a^3/4$ of the
FCC lattice we obtain
\begin{eqnarray} \label{k1k2}
\left[ (k_1/k_0)^2 - \bar{\varepsilon} \right] {\bm E}_1 =
\varepsilon' {\bm E}_2 + \sum_{j=1,2} {\bm E}_j \sum_{\nu}
I_{\nu}^{(1)*} I_{\nu}^{(j)} T_{\nu}\:,
\\
\left[ (k_2/k_0)^2 - \bar{\varepsilon} \right] {\bm E}_2 =
\varepsilon' {\bm E}_1 + \sum_{j=1,2} {\bm E}_j \sum_{\nu}
I_{\nu}^{(2)*} I_{\nu}^{(j)} T_{\nu} \:.\nonumber
\end{eqnarray}
The notation used are as follows: $k_0 = \omega/c$,
\[
\bar{\varepsilon} = \frac{1}{v_0}\int\limits_{v_0}\varepsilon({\bm
r})d^3r\:,\: \varepsilon' =\frac{1}{v_0}\int\limits_{v_0}e^{4 \pi
{\rm i} z/a} \varepsilon({\bm r}) d^3r \:,
\]
\begin{equation} \label{int}
I_{\nu}^{(j)} = \frac{1}{\sqrt{v_0}} \int\limits_{v_0}  e^{{\rm i}
k_j z} \Phi^*_{\nu}({\bm r}) d^3r\:,\: T_{\nu} =
\frac{\varepsilon_a \omega_{\rm LT}}{\omega_{\nu} - \omega}\:,
\end{equation}
and we assume the origin of the Cartesian coordinate system to be
chosen in the center of one of the spheres A.

In the absence of the exciton-photon coupling and dielectric
contrast, the right-hand sides of equations (\ref{k1k2}) vanish,
$\bar{\varepsilon} = \varepsilon_B$, and we arrive at two branches
of the dispersion relation (\ref{bare}) for ``bare'' photons. At
the $X$ point these branches converge. In the presence of
dielectric contrast, $\varepsilon' \neq 0$ and the fourfold
degenerate state at the $X$ point splits into doubly-degenerate
states $X_{5'}$ ($E_1 = E_2$) and $X_{5}$ ($E_1 = - E_2$) with
their eigen frequencies being
\begin{equation} \label{gap} \omega(X_5) =
\frac{c k_X}{\sqrt{ \bar{\varepsilon} - \varepsilon' } }\:,\:
\omega(X_{5'}) = \frac{c k_X}{ \sqrt{\bar{\varepsilon} +
\varepsilon' } }
\end{equation}
and the splitting being $\omega(X_5) - \omega(X_{5'}) \approx
(\varepsilon' / \bar{\varepsilon})(c k_X/\bar{n})$, where $k_X =2
\pi /a$, $\bar{n} = \sqrt{\bar{\varepsilon}}$. The approximate
equations (\ref{gap}) reproduce with high accuracy results of the
exact calculation presented in Fig.~4 by dashed lines.

In the absence of dielectric contrast, with allowance for the
exciton level $1s$ only and negligible difference between the
integrals $I_{1s}^{(1)}$ and $I_{1s}^{(2)}$ the set of equations
(\ref{k1k2}) reduces to
\begin{widetext}
\begin{equation} \label{2wave1s}
\left[ (k_1/k_0)^2 - \varepsilon_B \right] {\bm E}_1 = \left[
(k_2/k_0)^2 - \varepsilon_B \right] {\bm E}_2 = T_{1s} I_{1s}^2
({\bm E}_1 + {\bm E}_2) \:,
\end{equation}
\end{widetext}
where $I_{1s}$ is the integral (\ref{int}) corresponding to the
wave vector $k_X$. Provided that the frequencies $\omega_1$ and
$\omega_X \equiv c k_X/n_B$ are close to each other, the photonic
states $X_5$, $X_{5'}$ are replaced by three doubly-degenerate
polaritonic states: one of the symmetry $X_5$ with the frequency
$\omega_X$ and two of the symmetry $X_{5'}$ with the frequencies
\begin{equation} \label{32}
\omega = \frac{\omega_1 + \omega_X}{2} \pm
\sqrt{\left(\frac{\omega_1 - \omega_X}{2}\right)^2 + \delta^2}\:,
\end{equation}
where $\delta = \sqrt{\omega_1 \omega_{\rm LT} I_{1s}^2}$\ . For
coinciding frequencies $\omega_1$ and $\omega_X$, the polariton
spectrum in the direction ${\bm k} \parallel [001]$ has a
stop-band of the width $2 \delta$ centered at $\omega = \omega_1$.
The detuning of $\omega_X$ from $\omega_1$ leads to shifts of the
stop-band edges according to Eq.~(\ref{2wave1s}) and, in addition,
an allowed band is formed in the center of the stop-band,
similarly as it happens in the resonant Bragg quantum-well
structure (see \cite{voronov} and references therein).

For an approximate description of the solid lines in Fig.~4 it
suffices to keep in the sums over $\nu$ in Eqs.~(\ref{k1k2}) only
the contribution due to the exciton quantum-confined ground state.
In this approximation the mixing of the photon and exciton states
of the symmetry $X_{5'}$ leads to their repulsion and formation of
hybrid waves with the frequencies
\begin{equation} \label{32a}
\omega = \frac{\omega_1 + \omega(X_{5'})}{2} \pm
\sqrt{\left[\frac{\omega_1 - \omega(X_{5'})}{2}\right]^2 +
\delta^2}\:.
\end{equation}
The previous equation (\ref{32}) is a particular case of this more
general equation. Since the $1s$ quantum-confined exciton does not
interact with the light wave of the symmetry $X_5$, the $X_5$
photon frequency remains unchanged in the considered
approximation. This allows to understand the closeness of the
$X_5$-polariton frequencies calculated neglecting the exciton
effects and taking into account only the $1s$ level, see the
points of intersection of the lines 2 and 2$'$ with the vertical
line $X$. Allowance for the excited excitonic levels in
Eq.~(\ref{k1k2}) results in a shift of the frequency of the lowest
polariton branch $X_{5'}$ downwards (line 1$''$). However, their
influence is small in comparison with the $1s$ exciton. At the
same time, in the formation of the polariton $X_5$ (line 2$''$)
the main role is played by coupling of the photon $X_5$ with the
exciton $1p$ and $m=0$. The calculation shows that the branch
2$''$ is well described by the two-wave model (\ref{k1k2}) taking
into account only one excitonic state ($1p,m=0)$.
\section{Conclusion}
We have developed a theory of resonant three-dimensional photonic
crystals made up of two compositional materials A (spheres) and B
(dielectric matrix) for an arbitrary value of the exciton
effective mass $M$ and arbitrary dielectric contrast determined by
the difference between the dielectric susceptibility
$\varepsilon_B$ of the matrix and the background dielectric
constant $\varepsilon_a$  of the material A.

For a finite value of $M$, the lowest polariton branch lies
between the branches calculated in the two limiting cases, namely,
in the absence of space dispersion, i.e., for $M\to\infty$, and
with allowance for only one excitonic level, i.e., for $M\to0$ and
$\omega_1 = {\rm const}$. For a satisfactory description of the
polariton branches in the frequency region $\omega < \omega_1$, it
suffices to take into account the light coupling with few lowest
states of the mechanical exciton.

For a resonant photonic crystal with the dielectric contrast, we
have analyzed the exciton-induced modification of the stop-band in
the $[001]$ direction if the exciton resonance frequency
$\omega_1$ is chosen to lie in the middle of the stop-band of the
analogous photonic crystal without excitons. If the dielectric
contrast is weak so that $|\varepsilon_B - \varepsilon_a| \ll
\varepsilon_B$ then one can use the two-wave approximation that
allows describe with good accuracy the results of numerical
calculation. The main contribution into the position of the lower
edge of the polariton stop-band along the $\Gamma - X$ line comes
from the exciton $1s$ states interacting with the light wave of
the $X_{5'}$ symmetry. The upper edge of the stop-band is governed
by the coupling of the exciton $(1p, m=0)$ with the photon of the
$X_{5}$ symmetry.

It should be noted that the developed theory with $M\to\infty$ can
be applied for the study of photonic crystals infiltrated with dye
aggregates periodically distributed in the space and characterized
by a certain resonance frequency of optical transitions, see
\cite{raikh} and references therein. The theory can be applied as
well for the calculation of the dispersion of exciton-polaritons
in resonant two-dimensional photonic crystals, e.g., in periodic
arrays of cylinders A embedded into the matrix B.

The work is supported by the Russian Foundation for Basic Research
(grant 05-02-16372) and the program of the Russian Ministry of
Science and Education.

\end{document}